\documentclass[reprint,aps,physrev,superscriptaddress,floatfix]{revtex4-2}

\usepackage{amsmath,amssymb}
\usepackage{graphicx}
\usepackage{dcolumn}
\usepackage{bm}
\usepackage[normalem]{ulem}
\usepackage{lipsum} 
\usepackage{hyperref}
\usepackage{upgreek}
\hypersetup{
colorlinks=true, %colorise les liens
citecolor=magenta,
breaklinks=true, %permet le retour a la ligne dans les liens trop longs
urlcolor= blue, %couleur des hyperliens
linkcolor= blue, %couleur des liens internes
bookmarksopen=true, %si les signets Acrobat sont crees,
}
\newcommand{\eps}{\varepsilon}

\newcommand{\orcid}[1]{%
  \href{https://orcid.org/#1}{\raisebox{-0.05em}{\includegraphics[height=1em]{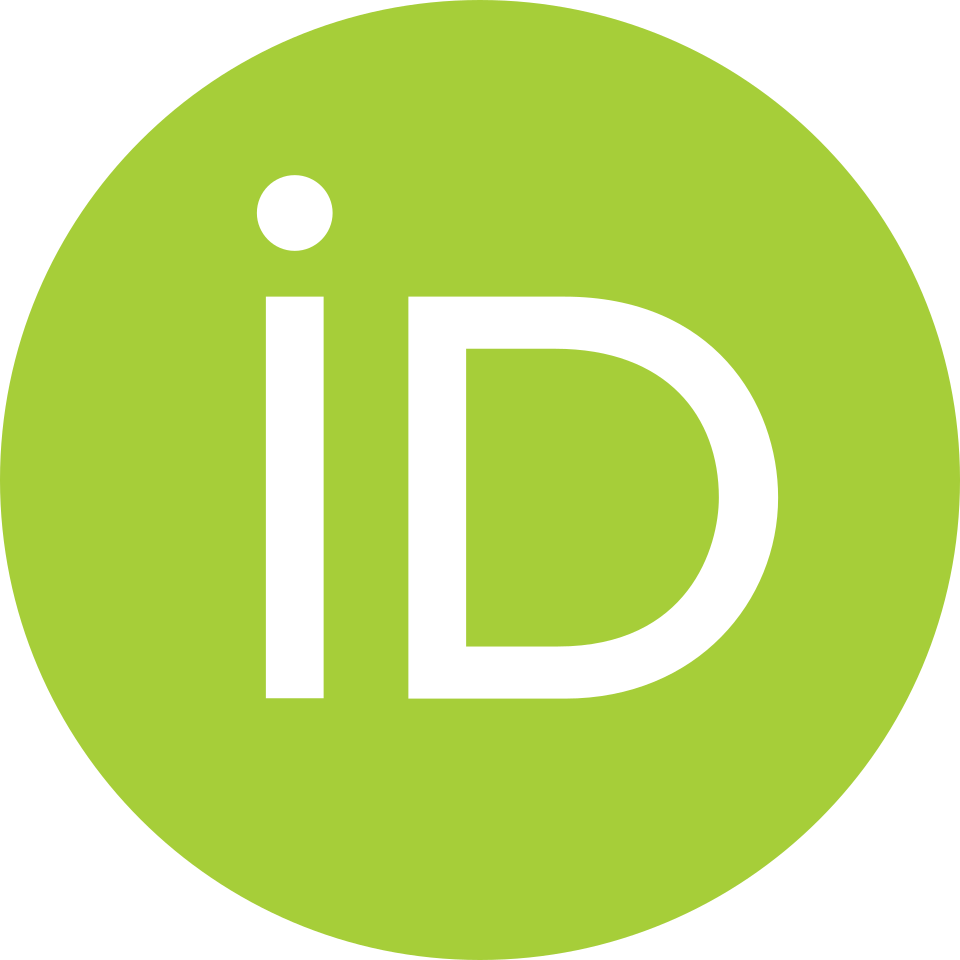}}}%
}

\begin{document}

\title{Fifth-order KdV solitary waves in a nonlinear electrical transmission line}

\author{Lo\"ic Fache\orcid{0009-0001-4172-5564}}
\affiliation{Universit\'e Paris Cit\'e, CNRS, MSC Laboratory, UMR 7057, F-75013 Paris, France}

\author{Filip Novkoski\orcid{0000-0003-0658-0336}}
\affiliation{PULS, Institute for Theoretical Physics, FAU Erlangen-N\"urnberg, 91058 Erlangen, Germany}

\author{Eric Falcon\orcid{0000-0001-9640-9895}}
\affiliation{Universit\'e Paris Cit\'e, CNRS, MSC Laboratory, UMR 7057, F-75013 Paris, France}

\date{\today}

\begin{abstract}
We report the controlled experimental observation of solitary waves governed by the fifth-order dispersion Korteweg-de Vries (KdV) equation. To achieve vanishing cubic dispersion, we use a mutual inductive coupling between nearest-neighbor inductors within a nonlinear electrical transmission line. We then experimentally characterize the resulting solitary waves through direct spatiotemporal measurements of the dispersion relation, its oscillatory tail profiles, amplitude-width scaling, and amplitude-velocity relation, finding good agreement with semi-analytical predictions. Collision experiments between two (co- or counter-propagating) solitary waves also show that their self-similar profile persists after interaction, up to a time shift. Dissipation effects are also quantified.
\end{abstract} 

\maketitle

\textit{Introduction.} Solitons are universal nonlinear excitations arising from a balance between nonlinearity and dispersion. They form a pillar of modern nonlinear and wave physics and are readily observed in a large range of domains, from fluids~\cite{russell1845report,korteweg1895xli,hammack1974korteweg,bourdin2010observation,ChabchoubPRL2013,CazaubielPRF2018,ricard2024soliton}, plasmas~\cite{ikezi1970formation}, granular matter~\cite{CostePRE1997}, and optical fibers~\cite{mollenauer1980experimental,KiblerNatPhys2010,copie2023space} to Bose-Einstein condensates~\cite{burger1999dark} and electrical lattices~\cite{hirota1970studies}. For integrable systems, solitons are fundamental nonlinear modes of certain nonlinear partial differential equations, such as the Korteweg-de Vries (KdV) equation, where their collisions are elastic, up to phase shifts~\cite{zabusky1965interaction,gardner1967method}. Integrability thus makes solitons more than mere localized pulses.

For nonintegrable systems, stable localized waves also exist but interact inelastically, with radiation emission and amplitude redistribution~\cite{gorshkov1981interactions,makhankov1978dynamics,kivshar1989dynamics,yang2010nonlinear}. Such coherent structures can be supported by the Kawahara equation (a generalized KdV model)~\cite{kawahara1972oscillatory,KhusnutdinovaPoF2018}. When the third-order dispersive term becomes small or vanishes, quadratic nonlinearity is balanced by the fifth-order dispersion, leading to the fifth-order Korteweg-de Vries (KdV5) equation~\cite{Remoissenet2013}. In contrast with the classical KdV soliton of cubic dispersion, the KdV5 solitary wave solutions do not admit a closed-form analytical expression and must, in general, be computed numerically~\cite{kawahara1972oscillatory}. Moreover, the fifth-order dispersive term alters the solitary-wave profile: asymptotic oscillatory tails are exhibited instead of a single-hump localized pulse for the KdV soliton. This feature plays an important role in their weak interactions and possible bound states~\cite{kawahara1972oscillatory,gorshkov1976interactions}.

In shallow-water capillary-gravity waves, the KdV5 regime is expected to occur near the transition from usual elevation to depression KdV solitons ~\cite{falcon2002observation,champneys2002true,hunter1983solitary,zufiria1987symmetry}. It should also appear when geometry and boundary conditions modify the effective dispersive balance~\cite{LeDoudicJFM2021}, as in a torus of fluid~\cite{novkoski2022experimental}. The KdV5 equation also governs weakly nonlinear shallow-water waves propagating on an ice-covered ocean surface~\cite{xia2002nonlinear}, and magnetized plasmas where the third-order dispersive term vanishes~\cite {kakutani1969weak}. However, such a fifth-order dispersive regime still remains elusive to achieve experimentally. Despite early experimental attempts in the 1970s~\cite{nagashima1979experiment,gorshkov1976interactions}, no systematic and quantitative characterization of the KdV5 solitary-wave profile, its scaling laws, or its resolved collision dynamics has been reported so far. Ref.~\cite{nagashima1979experiment} provides only a brief report, based on a single oscilloscope snapshot. It claims a linear amplitude-velocity scaling but provides no supporting data, nor any data on pulse-width scaling. It also states that pulses keep their identities after collisions, without presenting data. Ref.~\cite{gorshkov1976interactions} reports the formation of a soliton bound state and of a soliton lattice. 

Here, we report a full experimental characterization of KdV5 solitary waves. To do so, we exploit the tunability of a nonlinear electrical transmission line (NLTL) to achieve vanishing cubic dispersion. Direct spatiotemporal measurements first reveal the KdV5 dispersion relation. Then, individual KdV5 solitary waves are observed and quantified. Their oscillatory-tail profiles, amplitude-width relation, and amplitude-velocity relation are in good agreement with semi-analytical predictions of the KdV5 equation. Finally, we study collisions between two (co- or counter-propagating) solitary waves. 

\begin{figure}[t]
    \centering
    \includegraphics[width=\columnwidth]{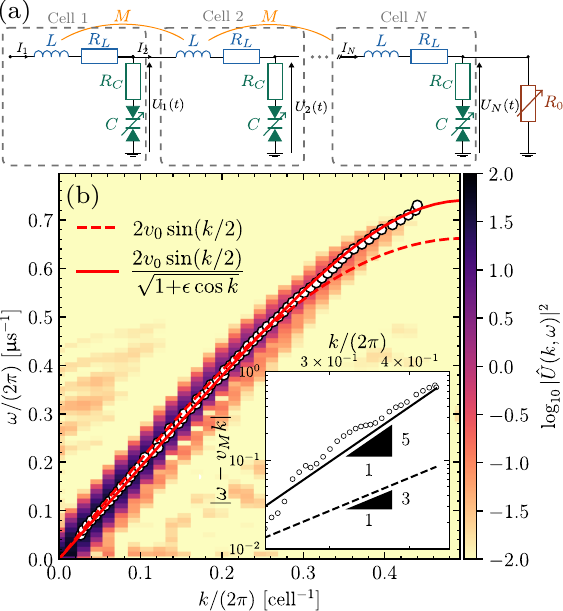}
    \caption{(a) Schematic of the nonlinear electrical lattice with $N = 64$ cells. Each cell contains a series inductance $L$ with internal resistance $R_L$, a varicap capacitance $C(U)$ with internal resistance $R_C$, and a mutual inductive coupling $M$ between nearest-neighbor inductors. A load resistance $R_0$ terminates the line. (b) Measured spatiotemporal power spectrum of the voltage field, $\log_{10}|\hat{U}(k,\omega)|^2$, in the linear regime. Crosses indicate the maxima of the experimental dispersion branch. Red lines are the theoretical discrete dispersion relation of Eq.~(\ref{eq:dispersion}): with $\eps=1/5$ (solid) and $\eps = 0$ (dashed). Inset: log-log zoom of the rescaled spectrum for $k/(2\pi) \in [0.25,\, 0.45]$.}
    \label{fig:setup}
\end{figure}

\textit{Experimental setup.} Our experimental platform is shown in Fig.~\ref{fig:setup}(a) and consists of a one-dimensional nonlinear electrical transmission line of $N = 64$ identical unit cells. Each cell $n$ contains a series inductance $L \simeq 330\,\upmu\mathrm{H}$ (with internal resistance $R_L \simeq 0.9~\Omega$) and a shunt nonlinear capacitance $C(U^{(n)})$ realized by a varicap diode (with internal resistance $R_C \simeq 12.8~\Omega$), whose capacitance decreases with the applied voltage $U^{(n)}$ and provides the required weak nonlinearity~\cite{Remoissenet2013}. The key ingredient is a mutual inductive coupling $M$ between nearest-neighbor inductors, thus distinguishing this device from a standard NLTL. The required value of $M$ is fixed by the intercoil spacing (with a center-to-center spacing of approximately $1.25~\mathrm{cm}$ in our setup). Input waveforms are generated using a programmable function generator (Rigol DG4102).

Applying Kirchhoff's voltage ($U$) and current ($I$) laws to cell $n$ of Fig.~\ref{fig:setup}(a), neglecting the internal series resistances $R_L$ and $R_C$, and accounting for the mutual flux threading each inductor, yields the coupled equations~\cite{nagashima1979experiment},
\begin{equation}
\small
\begin{aligned}
LI_t^{(n)}
+M\!\left[I_t^{(n-1)}+I_t^{(n+1)}\right]
&=U^{(n-1)}-U^{(n)},\\
C(U^{(n)}) U_t^{(n)}
&=I^{(n)}-I^{(n+1)},
\end{aligned}
\label{eq:discrete_nltl}
\end{equation}
where $t$ is the laboratory time, subscripts indicate derivatives, and the superscript $n$ labels the corresponding cell of the line. Experimentally, the varicap capacitance reads $C(U^{(n)})=C_0(1 - 2bU^{(n)})$, with $C_0 \equiv C(U=0)\simeq 683~\mathrm{pF}$ the zero-bias capacitance and $b \simeq 0.21~\mathrm{V^{-1}}$ a nonlinear coefficient independently calibrated from the capacitance-voltage characteristic (see Supp. Mat.~\cite{SuppMat}). The internal resistances $R_L$ and $R_C$, neglected in Eq.~\eqref{eq:discrete_nltl}, introduce distinct dissipative effects. In the long-wavelength limit, $R_L$ produces frequency-independent damping, whereas $R_C$ gives rise to a diffusion term associated with frequency-dependent dissipation (see Supp. Mat.~\cite{SuppMat} for details).

\textit{Dispersion Relation.} Seeking plane-wave solutions $U^{(n)} \propto e^{i(kn-\omega t)}$ in the linear approximation of Eqs.~\eqref{eq:discrete_nltl} gives the exact discrete dispersion relation $\omega(k)$ of the lattice~\cite{nagashima1979experiment}, as 
\begin{equation}
  \omega(k) = \frac{2}{\sqrt{LC_0}}\,
              \frac{\sin(k/2)}{\sqrt{1+\varepsilon\cos k}}\ \ \ {\rm with}
  \ \ \varepsilon \equiv \frac{2M}{L}.
  \label{eq:dispersion}
\end{equation}
For small wave numbers $k$, expanding Eq.~(\ref{eq:dispersion}) gives~\cite{nagashima1979experiment}
\begin{equation}
\small
\frac{\omega}{v_M} = k -\underbrace{\frac{1-5\varepsilon}{24(1+\varepsilon)}}_{\alpha_3}k^3 +\underbrace{\frac{1-58\varepsilon+121\varepsilon^2}{1920(1+\varepsilon)^2}}_{\alpha_5}k^5 +\cdots,
\label{eq:expansion}
\end{equation}
where $v_M = [(1+\varepsilon)LC_0]^{-1/2}$ is the long-wavelength phase speed in linear system. The mutual coupling $\varepsilon$ thus tunes the dispersive coefficients: the cubic coefficient $\alpha_3$ vanishes exactly at $\varepsilon = 1/5$, i.e., $M = L/10$, while the fifth-order coefficient satisfies $\alpha_5 < 0$ at this value. The leading dispersive correction thus scales as $k^5$, unlike in a standard NLTL ($\varepsilon \sim 0$).

First, we experimentally test the prediction of Eq.~\eqref{eq:dispersion} by measuring the linear impulse response of the line. We inject at the input of the NLTL a short voltage pulse $\delta U$ of duration $0.2\,\upmu\mathrm{s}$ and amplitude $10\,\mathrm{mV}$. The amplitude is small enough to ensure linear propagation, and the short duration provides a broadband excitation that probes, in a single measurement, the full band of propagating modes supported by the lattice. The voltage is then recorded at successive cells along the line to reconstruct the spatiotemporal field $U(n,t)$.

Then, by computing the two-dimensional Fourier power spectrum $|\hat{U}(k,\omega)|^2$, with $k$ the discrete wavenumber and $\omega$ the angular frequency, we obtain the experimental dispersion relation. As shown in Fig.~\ref{fig:setup}(b), the power concentrates along a main curve in the $(k,\omega)$ plane, which directly identifies the linearly propagating modes of the NLTL, in excellent agreement with the discrete dispersion relation of Eq.~(\ref{eq:dispersion}) evaluated at $\varepsilon=1/5$. By contrast, the theoretical dashed red curve obtained by setting $\varepsilon=0$, corresponding to a lattice without mutual inductive coupling, clearly departs from the experimental spectrum. The log-log inset further highlights this deviation by plotting $\omega-v_M k$ versus $k$. The data collapse well onto a $k^5$ power law, confirming the dominant fifth-order dispersive behavior.

\textit{Fifth-order KdV equation.} In the long-wavelength and weakly nonlinear approximations of Eq.~(\ref{eq:discrete_nltl}), and in the lossless limit, the lattice dynamics for the voltage field $U(n,t)$ reads (see Supp. Mat.~\cite{SuppMat})
\begin{equation}
  U_n + \frac{1}{v_M}U_t- \frac{b}{v_M} UU_t + \frac{1}{480 v_M^5}U_{ttttt} = 0.
  \label{eq:dimless}
\end{equation}
The first two terms describe linear wave propagation at speed $v_M$, while the last two terms provide the weak quadratic nonlinearity and the fifth-order dispersion. 
Our experiments span solitary-wave maximum amplitudes $U_{\mathrm{m}}\in [0.01,2.24]$~V and their durations~$\in[0.47,1.7]$~$\upmu$s, yielding weak nonlinear and dispersive parameters ($bU_{\mathrm{m}} \in [0.002,0.47]$) and ($1/[480v_M^{4}\tau^{4}]\in [0.003,0.44]$) both smaller than 1 and of the same order of magnitude as required for the KdV5 balance, with $\tau$ the dispersive time scale (see below and~\cite{footnot2}).

Moving to a reference frame co-propagating at $v_M$ and introducing the normalized variables $T = 480^{1/4}(bU_\mathrm{m})^{5/4}n$ and $x = (480b U_\mathrm{m})^{1/4}\left(n-v_Mt\right)$ together with the dimensionless voltage $u = U/U_\mathrm{m}$, Eq.~(\ref{eq:dimless}) then reduces to the dimensionless KdV5 equation
\begin{equation}
  u_T + uu_x - u_{xxxxx} = 0.
  \label{eq:KdV5}
\end{equation}

\begin{figure}[t]
    \centering
    \includegraphics[width=\columnwidth]{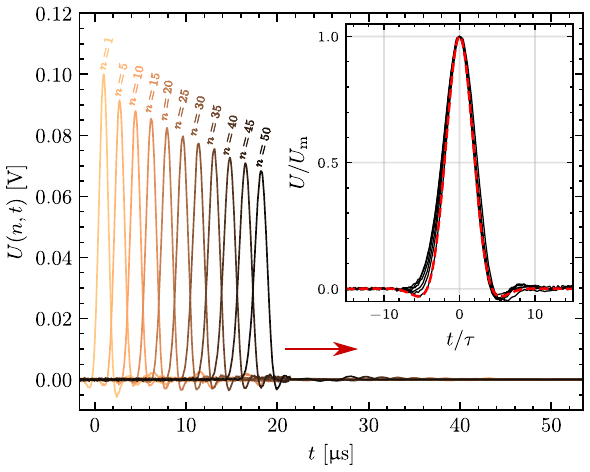}
    \caption{Propagation of an injected KdV5 solitary-wave profile. Main: voltage profiles $U(n,t)$ at cell indices $n$, from $n=1$ (light orange) to $n=50$ (black), as a function of laboratory time $t$.  Inset: rescaled profiles $U/U_\mathrm{m}$ vs. $t/\tau$ for several selected cells (solid lines) collapse on the normalized KdV5 solitary-wave solution of Eq.~\eqref{eq:KdV5} (red-dashed line). }
    \label{fig:solitary}
\end{figure}
\textit{Solitary wave: profile and velocity.} Unlike the classical KdV equation, Eq.~(\ref{eq:KdV5}) does not admit closed-form solitary-wave solutions. The fifth-order dispersive term generically produces oscillatory tails that prevent the simple hyperbolic-secant ansatz from closing, and no exact localized solution has been found in the general case~\cite{kawahara1972oscillatory}. Localized traveling-wave solutions can nevertheless be obtained numerically using the Petviashvili fixed-point iteration in Fourier space~\cite{petviashvili1976equation,yang2010nonlinear} (see Supp. Mat.~\cite{SuppMat}). The iteration converges to a localized profile featuring a positive main crest and exponentially damped oscillatory tails of alternating sign. In the experiment, this numerical profile is mapped back to a physical voltage waveform through the above scaling relations and injected as the input voltage $U(n=0)$.

Figure~\ref{fig:solitary} shows the propagation of a single injected pulse. The voltage profiles $U(n,t)$ recorded at successive cell indices $n$ (color-coded from light orange to black) exhibit a localized waveform traveling at constant speed while preserving its self-similar shape over the full propagation range (see inset). The slight deviations are due to dissipative effects. A key signature of fifth-order dispersion is the amplitude-duration scaling of the solitary wave, which differs fundamentally from the KdV case. For the standard KdV equation, the soliton duration scales as $\tau \approx 1/[v_0(24\,bU_\mathrm{m})^{1/2}]$~\cite{Remoissenet2013}, $U_\mathrm{m}$ being its maximum amplitude, reflecting the balance between quadratic nonlinearity and cubic dispersion. For Eq.~\eqref{eq:dimless}, the balance between the same nonlinearity and fifth-order dispersion instead imposes $\tau \approx 1/[v_M (480bU_\mathrm{m})^{1/4}]$. When the measured profiles are rescaled as $U/U_\mathrm{m}$ and plotted against $t/\tau$ [inset of Fig.~\ref{fig:solitary}], waveforms then collapse onto a single master curve, which superimposes with the numerical solution of Eq.~\eqref{eq:KdV5} (red-dashed line). Note that the experimental width, which is slightly greater than the latter, is due to dissipation effects. Note also that the profile exhibits troughs on either side of the hump, which is a signature of a KdV5 profile. The sensitivity of this evolution to the injected waveform is further investigated in the Appendix, where localized inputs of different widths are used to probe pulse propagation, lattice discreteness, and dissipation effects [Fig.~\ref{fig:width_sweep}].

\begin{figure}[t]
    \centering
    \includegraphics[width=\columnwidth]{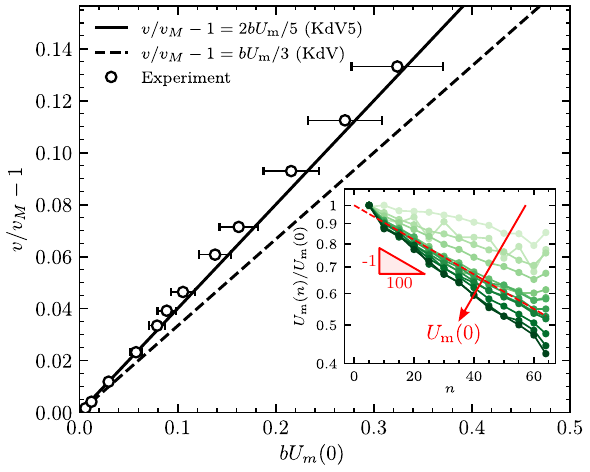}
    \caption{ Velocity-amplitude scaling of KdV5 solitary waves. The measured normalized velocity $v/v_M-1$ is plotted versus $bU_{\mathrm m}(0)$ (circles). Solid line: KdV5 prediction of Eq.~\eqref{eq:velocity_sol_kdv5}. Dashed line: standard KdV prediction. Horizontal error bars reflect the amplitude variation during propagation. Inset: Normalized peak voltage $U_{\mathrm m}(n)/U_{\mathrm m}(0)$ along the NLTL for various input $U_{\mathrm m}(0)$ amplitudes in semi-log-$y$.}
    \label{fig:scaling}
\end{figure}
The scaling of the measured velocity of the pulse with its amplitude is also directly compared with the prediction of the reduced KdV5 dynamics. Using the rescaled variables and $x-c_\mathrm{KdV5}T = \mathrm{const.}$, with $c_\mathrm{KdV5} = 2/5$~\cite{yamamoto1981solution}, the solitary-wave velocity is predicted to scale linearly with its amplitude, as
\begin{equation}\label{eq:velocity_sol_kdv5}
    v = v_M \left(1 + \frac{2}{5} b U_{\mathrm m}(0) \right).
\end{equation}
The experimental pulse velocity is inferred from the time of flight between successive waveform maxima across all NLTL cells. The resulting measurements, reported in Fig.~\ref{fig:scaling}, are in good agreement with Eq.~\eqref{eq:velocity_sol_kdv5} (see solid line). Note that in the case of standard KdV dynamics, the scaling predicts $v = v_0 \left(1 + b U_\mathrm{m}/3 \right)$~\cite{Remoissenet2013} (see dashed line). The inset of Fig.~\ref{fig:scaling} shows the system dissipation: in the worst case, the solitary-wave amplitude decays exponentially with a rate of about 10$^{-2}$ cell$^{-1}$. This attenuation mainly originates from the internal series resistances $R_L$ and $R_C$ of the inductors and varicap diodes, respectively (see Supp. Mat.~\cite{SuppMat} for details). 

\textit{Soliton collisions.} A hallmark of solitary waves is their ability to interact while preserving their identities after the collision. We test this property in the KdV5 regime through head-on and overtaking collisions between two counter- or co-propagating solitary structures.
\begin{figure}[t]
    \centering
    \includegraphics[width=\columnwidth]{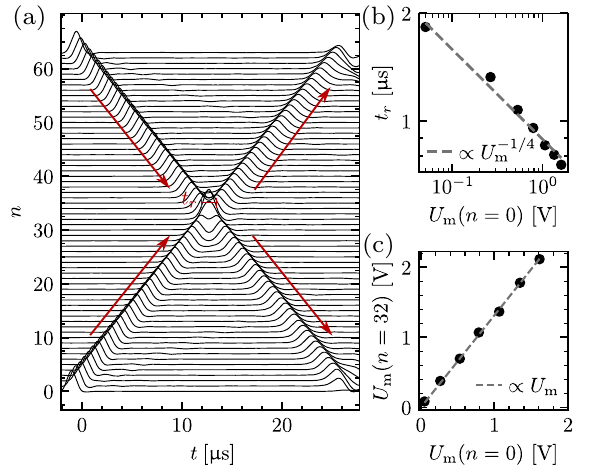}
    \caption{Head-on collision between two KdV5 solitary waves in the NLTL. (a) Spatiotemporal evolution of the voltage field $U(n,t)$ during the collision. The two counter-propagating solitary waves are generated with identical input parameters, with peak amplitude $0.26~\mathrm{V}$ and duration $0.84~\mathrm{\upmu s}$. (b) The residence time scales as $t_r\propto U_\mathrm{m}(n=0)^{-1/4}$ (Log-Log). (c) Collision peak amplitude (at $n=32$)  is proportional to the injected one (with a slope of $\sim 1.3$). (b,c) Amplitude sweep forcing. Dashed lines correspond to power-law scalings.}
    \label{fig:collision}
\end{figure}

Figure~\ref{fig:collision}(a) shows a typical head-on collision experiment. Two KdV5 solitary waves are launched from opposite ends of the NLTL and collide near the center of the lattice. The trajectories before and after interaction show that both waves recover nearly unchanged profiles and velocities after collision. The interaction region is characterized by a transient nonlinear superposition accompanied by a localized increase in amplitude. Equation~(\ref{eq:KdV5}) is a unidirectional model, whereas the NLTL supports waves propagating in both directions. A bidirectional long-wave description of NLTLs is more naturally formulated at the level of Boussinesq-type equations, before projection onto a single propagation direction~\cite{Remoissenet2013}.

Using the concept of residence time~\cite{power1984reflection,novkoski2022experimental}, we characterize the local duration of the collision by the time $t_r$ during which the spatial profile $U(n)$ cannot be distinguished into two distinct maxima. Since the KdV5 balance gives a characteristic temporal width scaling as $U_\mathrm{m}^{-1/4}$, instead of the classical KdV scaling $U_\mathrm{m}^{-1/2}$~\cite{Remoissenet2013}, one expects $t_r \propto U_\mathrm{m}^{-1/4}$. This scaling is consistent with the experiments reported in Fig.~\ref{fig:collision}(b). Fig.~\ref{fig:collision}(c) shows that the peak voltage measured at the collision location, $n=32$, remains proportional to the injected amplitude.

Unlike classical solitons of the integrable KdV equation, KdV5 solitary waves are not expected to collide elastically, as shown in numerical studies~\cite{malomed1996solitary}. Here, the localized structures persist after interaction, with weak disturbances attributable to both collision dynamics and unavoidable NLTL dissipation. However, these two contributions cannot be distinguished separately in our experiments.

\begin{figure}[t]
    \centering
    \includegraphics[width=\columnwidth]{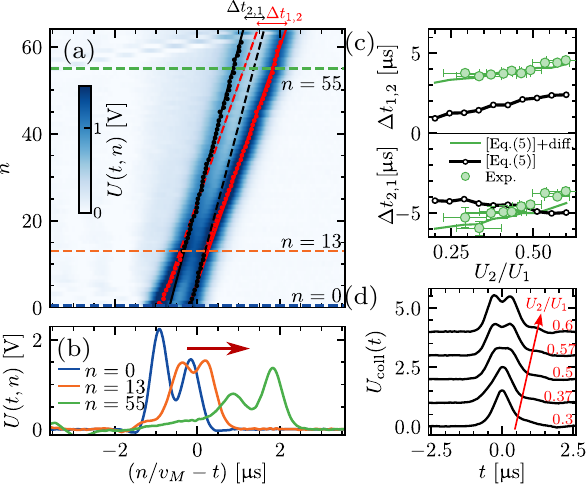}
    \caption{Overtaking collision between two KdV5 solitary waves. (a)~Spatiotemporal evolution of the voltage field $U(n,t)$ in the frame moving at $v_M$.  (b)~Temporal signals measured at cells $n=0$ (blue), 13 (orange), and 55 (green). The initial large (small) solitary wave has a peak amplitude of $2.24~\mathrm{V}$ ($1.56~\mathrm{V}$) and a duration of $0.47~\mathrm{\upmu s}$ ($0.52~\mathrm{\upmu s}$). (c) Collision-induced time shifts of the large ($\Delta t_{1,2}>0$) and small ($\Delta t_{2,1}<0$) waves versus their amplitude ratio $U_2/U_1$. Green circles: experiments. Black piecewise line: KdV5 simulations. Green solid line: simulations of the dissipative KdV5 equation. (d) Profiles at maximum interaction for different $U_2/U_1$, vertically shifted for clarity.}
    \label{fig:overtaking}
\end{figure}

Figure~\ref{fig:overtaking}(a,b) shows an overtaking collision between two co-propagating KdV5 solitary waves of different amplitudes. Consistent with the velocity scaling reported in Fig.~\ref{fig:scaling}, the larger-amplitude wave propagates faster, catches up with the smaller one, and overtakes it. During the interaction, the two localized structures transiently merge before separating again into two solitary-wave packets [see Fig.~\ref{fig:overtaking}(a,b)].

The interaction produces a clear collision-induced time shift: the larger wave (No.~1) is advanced, whereas the smaller one (No.~2) is delayed relative to their unperturbed trajectories. Figure~\ref{fig:overtaking}(c) shows that the measured shifts $\Delta t_{1,2}$ and $\Delta t_{2,1}$ follow, for various amplitude ratio $U_2/U_1$, the same overall trend as numerical simulations of Eq.~\eqref{eq:KdV5}  (black line). Moreover, $\Delta t_{1,2}<|\Delta t_{2,1}|$ as the smaller pulse (i.e., the slowest one) interacts longer. The departure from Eq.~\eqref{eq:KdV5} is ascribed to dissipation in the experimental line, which changes the solitary-wave amplitudes and velocities during propagation, as captured by simulations of the dissipative KdV5 equation (see Supp. Mat.~\cite{SuppMat}). Finally, Fig.~\ref{fig:overtaking}(d) shows the soliton-interaction profile for different pulse-amplitude ratios. 

\textit{Conclusion.} We have demonstrated that a nonlinear electrical transmission line with a mutual inductive coupling realizes an experimental regime in which the cubic term in the long-wavelength expansion of the lattice dispersion relation vanishes, leaving fifth-order dispersion as the leading dispersive correction. Using spatiotemporal measurements, we have experimentally highlighted the observation of KdV5 solitary waves in good agreement with the predictions: oscillatory-tail profile, amplitude-width scaling, and linear amplitude-velocity relation. Collisions between two (co- and counter-propagating) solitary waves have also been investigated. In future, we could explore  another solitary-wave collision type in which only interaction occurs through their oscillatory tails~\cite{Nagashima1981}. Novel KdV5 solutions as dispersive shock waves could also be investigated in our NLTL~\cite{Sprenger2020}.

Beyond isolated collisions, future experiments could be tested using a large ensemble of such solitary waves, namely \textit{nonintegrable soliton gases}~\cite{flamarion2024non}, likely requiring lower-loss implementations. While soliton gas experiments have been performed so far in nearly integrable systems (such as in hydrodynamics~\cite{redor2019experimental}, optics~\cite{suret2023soliton}, and electrical lattices~\cite{fache2025dissipation}), the KdV5 regime provides an accessible nonintegrable counterpart. It thus raises the question of whether an effective kinetic theory, as the one derived for integrable systems~\cite{zakharov1971kinetic}, can be developed. Another intriguing open question is whether nonlinear discrete chains such as the NLTL can exhibit a statistical equilibrium regime~\cite{Onorato2023}, as recently observed for the large scales in three-dimensional turbulence~\cite{3DTurbES} and wave turbulence~\cite{WTES}. Finally, our results pave the way for observing KdV5 solitary waves in diverse physical systems, including internal waves~\cite{KhusnutdinovaPoF2018}, ice-covered oceans~\cite{xia2002nonlinear}, and magnetized plasmas~\cite{kakutani1969weak}, where their detection remains an open challenge.

\begin{acknowledgments}
\textit{Acknowledgments.} We thank C.-T. Pham for discussions. This work was supported by the Simons Foundation Project No. MPS-WT-00651463 (U.S.) on Wave Turbulence and the French National Research Agency (ANR Sogood Project No.~ANR-21-CE30-0061-04, ANR Lascaturb Project No.~ANR-23-CE30-0043-02, and ANR Provebact ANR-24-CE09-1394-02). F.N. thanks the Alexander von Humboldt Foundation for a postdoctoral fellowship.
\end{acknowledgments}

\textit{Data availability.} The data that support the findings of this article are not publicly available. The data are available from the authors upon reasonable request.

\section*{Appendix}  %(limited to 2 pages)
\setcounter{figure}{0} % Réinitialise le compteur des figures
\renewcommand{\thefigure}{A\arabic{figure}} % Renomme le préfixe des figures en "EM"
\textit{Evolution of non-solitonic localized inputs.---} To test the sensitivity of the dynamics to the injected waveform, we performed experiments at fixed peak voltage $U_m(n=0)$ using either the stationary KdV5 solitary-wave profile or Gaussian pulses of full width at half maximum (FWHM) $W/2$, $W$, $2W$, and $4W$, where $W\simeq0.96~\upmu\mathrm{s}$ is the solitary-wave FWHM [Fig.~\ref{fig:width_sweep}(a1)--(a5)]. The first column, highlighted by a transparent cyan mask, corresponds to the KdV5 solitary-wave case investigated in the main text and serves here as the reference. The corresponding spatial widths are approximately $0.9$, $1.9$, $3.7$, and $7.4$ cells, respectively.

The experiments reveal three distinct behaviors. The Gaussian pulse with the same amplitude and width as the KdV5 solitary wave evolves toward a waveform very close to that obtained from the exact solitary-wave input [see (a1) and (a3)], showing that propagation at this scale is not critically sensitive to the detailed input shape. By contrast, the $W/2$ pulse broadens strongly and develops an oscillatory tail, while retaining a localized core after $n=40$ cells [see (a2)], indicating the presence of lattice dispersive radiative components. For the broader $2W$ and $4W$ inputs, their profiles steepen and break into a dominant localized pulse accompanied by secondary structures, reminiscent of soliton fission in KdV dynamics~\cite{zabusky1965interaction,Remoissenet2013}.

The corresponding spatiotemporal Fourier spectra in Figs.~\ref{fig:width_sweep}(b1)--(b5) clarify these observations. A coherent structure propagating approximately at constant velocity produces spectral energy along a straight branch $\omega=vk$, whereas dispersive radiation follows the exact discrete linear dispersion relation, Eq.~\eqref{eq:dispersion}. The KdV5 input and the Gaussian pulse of width $W$ display very similar coherent branches, with only weak dispersive radiation [see (b1) and (b3)]. For the $W/2$ input, however, the initial spectrum extends to large wavenumbers, and a clear component develops along the discrete linear dispersion curve, directly revealing the emission of lattice radiation [see (b2)]. The surviving localized component is associated with a separate coherent branch. The finite propagation distance prevents us from establishing asymptotic convergence toward an exact solitary wave, but the measurements clearly show a nonlinear readjustment accompanied by dispersive-wave emission.

Finally, Fig.~\ref{fig:width_sweep}(c1)--(c5) shows the corresponding simulations of the discrete circuit equations with the resistive components removed [i.e., Eq.~\eqref{eq:discrete_nltl}]. Comparison with experiments and dissipative simulations in (a1)--(a5) isolates the role of the internal resistances $R_C$ and $R_L$. In the lossless lattice, dispersive and secondary structures persist more strongly, whereas dissipation attenuates both the coherent pulses and the emitted radiation.

\begin{figure*}[t!]
\includegraphics[width=\textwidth]{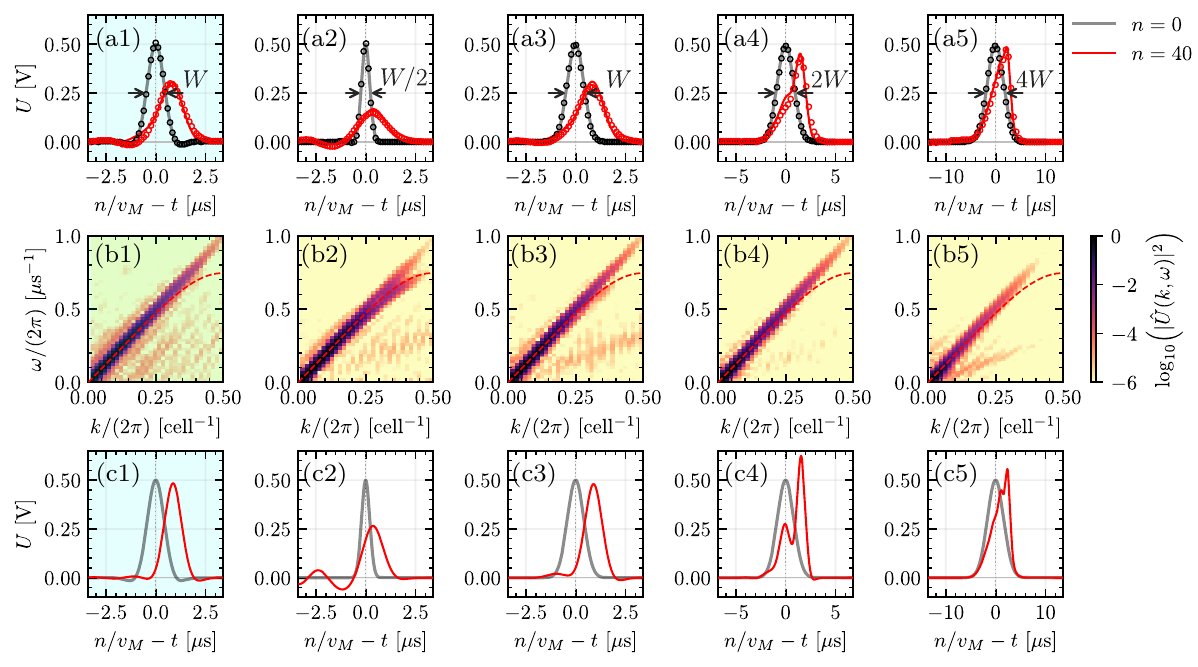} 
\caption{Evolutions of different localized-wave profiles injected at $n=0$ (grey) propagating along the NLTL at  $n=40$ (red). Columns show, from left to right, a KdV5 solitary wave of FWHM $W$ and Gaussian pulses of FWHM $W/2$, $W$, $2W$, and $4W$, for fixed peak voltage $U_\mathrm{m}(n=0) \simeq 0.5~\mathrm{V}$. $W\simeq0.96~\upmu\mathrm{s}$. (a1--a5) Experiments (open circles) along with the corresponding dissipative discrete-NLTL simulations (solid lines). (b1--b5) Experimental spatiotemporal Fourier spectra, along with the exact discrete linear dispersion relation [Eq.~\eqref{eq:dispersion} - red-dashed line]. (c1--c5) Lossless simulations of the discrete-NLTL Eq.~\eqref{eq:discrete_nltl}.}
\label{fig:width_sweep}
\end{figure*}

\end{document}